\documentclass[12pt]{JHEP3}

\usepackage{amssymb}
\usepackage{graphicx}
\usepackage{amsmath}
\usepackage{epsfig}
\usepackage{amsfonts}

\newcommand{\beq}{\begin{equation}}
\newcommand{\eeq}{\end{equation}}
\newcommand{\ben}{\begin{enumerate}}
\newcommand{\een}{\end{enumerate}}
\newcommand{\bit}{\begin{itemize}}
\newcommand{\eit}{\end{itemize}}
\newcommand{\beqray}{\begin{eqnarray*}}
\newcommand{\eeqray}{\end{eqnarray*}}

\newcommand{\hmpcinv}{h \mathrm{Mpc}^{-1}}

\newcommand{\Omegam}{\Omega_{\mathrm{m}}}
\newcommand{\Omegab}{\Omega_{\mathrm{b}}}

\newcommand{\ns}{n_{\mathrm{s}}}
\newcommand{\dr}{\Delta_{\mathcal{R}}^{2}}
\newcommand{\da}{\mathrm{D_A}}
\newcommand{\daz}{\mathrm{D_A}(z)}

\newcommand{\dasqz}{\mathrm{D}_{\mathrm{A}}^{2}(z)}

\newcommand{\lcdm}{\Lambda\mathrm{CDM}}

\newcommand{\wzero}{w_{0}}
\newcommand{\wa}{w_{\mathrm{a}}}

\newcommand{\kappai}{\kappa_{\mathrm{i}}}
\newcommand{\kappaj}{\kappa_{\mathrm{j}}}

\newcommand{\gi}{g_{\mathrm{i}}}
\newcommand{\gj}{g_{\mathrm{j}}}

\newcommand{\pkikj}{P^{\kappai\kappaj}}
\newcommand{\pgigj}{P^{\gi\gj}}
\newcommand{\pkigj}{P^{\kappai\gj}}
\newcommand{\pxixj}{P^{\xsubi\xsubj}}
\newcommand{\pxmxn}{P^{\xsubm\xsubn}}

\newcommand{\wiz}{\mathrm{W}_{\mathrm{i}}(z)}
\newcommand{\wjz}{\mathrm{W}_{\mathrm{j}}(z)}

\newcommand{\ssubi}{\mathrm{s}_{\mathrm{i}}}
\newcommand{\ssubj}{\mathrm{s}_{\mathrm{j}}}

\newcommand{\xsubi}{\mathrm{x}_{\mathrm{i}}}
\newcommand{\xsubj}{\mathrm{x}_{\mathrm{j}}}

\newcommand{\xsubm}{\mathrm{x}_{\mathrm{m}}}
\newcommand{\xsubn}{\mathrm{x}_{\mathrm{n}}}

\newcommand{\deltasubi}{\delta_{\mathrm{i}}}
\newcommand{\deltasubj}{\delta_{\mathrm{j}}}

\newcommand{\niz}{n_\mathrm{{i}}(z)}
\newcommand{\zphot}{z^{\mathrm{ph}}}
\newcommand{\phzpdf}{P(\zphot|z)}

\newcommand{\zbias}{z_{\mathrm{bias}}}

\newcommand{\na}{N_{\mathrm{A}}}

\newcommand{\galarcmintwo}{\mathrm{ gal / arcmin}^{2}}

\newcommand{\pmat}{\mathrm{P}_{\delta}}

\newcommand{\fsky}{f_{\mathrm{sky}}}

\newcommand{\dd}{\mathrm{d}}

\newcommand{\sonehalf}{S_{1/2}}
\newcommand{\logten}{\mathrm{log}_{10}}
\newcommand{\logtenkc}{\mathrm{log}_{10}({k_{c}}/\hmpcinv)}

\title{Testing the Origin of the CMB Large-Angle Correlation Deficit with a Galaxy Imaging Survey}

\author{Andrew P. Hearin\footnotemark[1], 
Cameron Gibelyou\footnotemark[2], 
and Andrew R. Zentner\footnotemark[1]\\
$^*$Pittsburgh Particle Astrophysics and Cosmology Center \& \\
Department of Physics and Astronomy, University of Pittsburgh,\\ 
Pittsburgh, PA 15260 USA\\
$^{\dagger}$Department of Physics, University of Michigan, \\
Ann Arbor, MI 48109 USA
}

\abstract{
The cosmic microwave background (CMB) temperature 
distribution measured by the Wilkinson Microwave Anisotropy Probe 
(WMAP) exhibits anomalously low correlation at large angles. 
Quantifying the degree to which this feature in the temperature data is in conflict 
with standard $\lcdm$ cosmology is somewhat ambiguous because of the 
\textit{a posteriori} nature of the observation.  One physical mechanism that has been 
proposed as a possible explanation for the deficit in the large-angle temperature correlations is 
a suppression of primordial power on $\sim$~Gpc scales. 
To distinguish whether the anomaly is a signal of new physics, such as suppressed primordial power, 
it would be invaluable to perform experimental tests of the authenticity of this signal in data sets which are 
independent of the WMAP temperature measurements or even other CMB measurements.  
We explore the possibility of testing models of power suppression with large-scale structure observations, 
and compare the ability of planned photometric and spectroscopic surveys to constrain the power spectrum. 
Of the surveys planned for the next decade, a spectroscopic redshift survey such as BigBOSS will have a 
greater number of radial modes available for study, but we find that this advantage is outweighed by the greater 
surface density of high-redshift sources that will be observed by photometric surveys such as LSST or Euclid.  
We also find that the ability to constrain primordial power suppression is insensitive to the precision of the calibration of photometric redshifts.
We conclude that very-wide-area imaging surveys have the potential to probe viable models for the missing power 
but that it will be difficult to use such surveys to conclusively rule out primordial power suppression as 
the mechanism behind the observed anomaly. 
}


\begin{document}

\section{Introduction}
\label{section:intro}

The consistency of the $\lcdm$ model of cosmology with the cosmic microwave
background (CMB) data observed by the Wilkinson Microwave Anisotropy Probe (WMAP) 
is one of the crowning achievements of twentieth-century cosmology.
Indeed, these observations were among the key results that led to the widespread acceptance
of the ``concordance model'' of cosmology.  

Despite the remarkable
agreement between $\lcdm$ predictions and the WMAP data, several anomalies on the largest
angular scales have persisted 
(see Refs.~\cite{copi_etal10} and \cite{bennett_etal11} for recent reviews).
Arguably the most troubling anomaly is the near-total lack
of correlation in the temperature anisotropy distribution on large scales: the CMB temperature autocorrelation function $C(\theta)$ as measured by WMAP is near
zero for angular scales above 60 degrees.  This puzzling observation first appeared in
the Cosmic Background Explorer (COBE) data \cite{smoot_etal92}
 before detection at much higher significance by WMAP. 
 Quantifying how unusual the large-angle correlation deficit is must be done with care because none of the statistics describing the anomaly itself were among the estimators proposed by the WMAP team prior to undertaking their analysis; that is, analysis of the correlation deficit is complicated by its {\em a posteriori} observation.
Recent estimates of the degree to which this anomaly is in conflict with $\lcdm$ vary significantly 
(e.g.,~Refs.~\cite{copi_etal07,copi_etal09, pontzen_peiris10,efstathiou_etal10}) 
and especially depend on the treatment of the Galactic region of the microwave sky. 
For example, the analysis in Ref.~\cite{efstathiou_etal10} employs a reconstruction technique to first generate a full-sky map and then use an all-sky estimator of $\mathcal{C}(\theta),$ concluding that the lack of CMB temperature correlation at large angles is unlikely at roughly the $95\%$ level. 
However, in a recent analysis \cite{copi_etal11} it was demonstrated that leakage of information from the masked region of the sky can lead to biases in the low multipoles computed from a reconstructed full-sky map (see also Ref.~\cite{feeney_etal11}). 
In an alternative approach to Ref.~\cite{efstathiou_etal10}, the analysis in Ref.~\cite{copi_etal09} uses a pixel-based estimator of $\mathcal{C}(\theta)$ that is constructed strictly from a cut-sky map to conclude that the possibility of the large-angle correlation deficit being a statistical fluke is unlikely at the level of $99.975\%.$   
The appropriate treatment of the Galactic plane remains an active area of research and a widespread consensus has not yet been reached, though this issue is central to the analysis of the low-multipole anomalies. 
Nonetheless, it is quite clear that the correlation deficit is not likely to be explained away as a simple systematic error:
as demonstrated in Ref.~\cite{bunn_bourdon08}, this feature cannot be accounted for
with a statistically independent contaminant such as an undiagnosed foreground, as such a contaminant
would only contribute additional large-scale power, thereby exacerbating the anomaly.

One possible explanation that would naturally produce the observed deficit in the correlation function at large angles is
that the primordial power spectrum generated in the early universe is suppressed 
on comoving scales comparable to the size of the horizon at the time of recombination.  
The persistence of the large-angle anomaly, coupled with the possibility of accounting for it with suppressed 
primordial power, has motivated many studies of possible mechanisms for the suppression in the context of inflation,
 e.g., Refs.~\cite{jing_fang94,linde03,contaldi_etal03,kuhnel_schwarz10,piao_etal04,lasenby_doran05,bridle_etal03,boyanovsky_etal06,feng_zhang03,yokoyama99,sinha_souradeep06,jain_etal09}.  
Thus the deficit in $C(\theta)$ at large angles may be an
indication that the simplest and most widely accepted models of inflation require revision.  

The distinct advantage of using the CMB to probe very large scales
is that the photons which free stream to Earth from the surface of last scattering have the potential to transmit information to us from the highest redshift in the visible universe.
Thus, in addition to CMB temperature, the polarization signal in the CMB can also potentially be exploited to provide useful information 
about the distribution of matter on the largest scales ~\cite{mortonson_hu09,mortonson_dvorkin_etal09}. 
Additionally, the scattering of CMB photons off of galaxy clusters induces a polarization signal that may be possible 
to exploit to probe very large scales~\cite{baumann_cooray03,bunn06,kamionkowski_loeb97,seto_pierpaoli05}. 
However, if primordial power in the early universe was in fact suppressed on large scales, then the signature
of this suppression should in principle also be imprinted in the distribution of large-scale structure at low redshift.
Confirmation of this signal in, for example, galaxy clustering statistics
would independently test the authenticity of the
low-power anomaly as a genuine feature of our cosmology.

In Ref.~\cite{gibelyou10}, the authors demonstrate that a forthcoming redshift survey such as BigBOSS 
\cite{bigboss09} has the potential to constrain viable models of primordial power suppression (see also Ref.~\cite{kesden_etal03}).
In this work, we attempt to answer a related question: what is the potential for future galaxy imaging surveys 
 such as the Large Synoptic Survey Telescope (LSST) \cite{lsst_sciencebook}
 or Euclid \cite{euclid_sciencebook} to test 
 the authenticity of the large-angle CMB temperature correlation deficit?
 On the one hand, photometric redshift (hereafter {\em photo-z}) uncertainty restricts the number of radial modes that will be available to an imaging survey relative to a data set with spectroscopic redshifts.  
 However, the much larger surface density of sources in an imaging survey dramatically reduces errors due to shot noise at high redshift and allows for the possibility of utilizing the cosmic shear signal in addition to galaxy clustering. 
 Thus through a joint analysis of cosmic shear and galaxy clustering, an imaging survey such as LSST or Euclid 
 may be able to provide independent tests of the large-angle CMB anomalies in the near future. 
 
This paper is organized as follows.  In \S\ref{section:methods} we describe our methods for modeling 
primordial power suppression and assessing its detectability in the distribution of large-scale structure.
In \S\ref{section:results} we present our results, and we discuss our conclusions in \S\ref{section:discussion}.

\section{Methods}
\label{section:methods}

\subsection{Primordial Power Suppression}
\label{sub:suppression}

To model the suppression of power on large scales, 
we modify the dimensionless curvature power spectrum 
$\Delta_{R}^2 (k) \equiv k^3 P_{R}(k) / 2 \pi^2$ 
by introducing a prefactor that encodes the exponential suppression, 
$\Delta^{2}_{R}(k)\rightarrow\Delta^{2}_{R}(k)\mathcal{S}(k),$ with 

\beq
\label{eq:pksupp}
\mathcal{S}(k)\equiv1-\beta\exp\left[-(k/k_{c})^{\alpha}\right].
\eeq
In Eq.~\ref{eq:pksupp}, $k_c$ governs the comoving scale at which 
suppression in $\Delta^{2}_{R}$ becomes significant,
$\beta$ the maximum fractional amount of suppression, and $\alpha$ the rapidity with which the suppression approaches
its maximal effect.  
We model $\Delta^{2}_{R}$ as in Eq.~\ref{eq:pksupp} not to advocate a particular alternative physical theory of the early universe, but rather so that we can have a simple model for the deficit in $C(\theta)$ at large angles whose consequences can then be explored.  
As for the particular functional form we choose for $\mathcal{S}(k),$ our motivation is twofold. 
First, the authors in Ref.~\cite{gibelyou10} demonstrated that $\lcdm$ cosmology with appropriately tuned exponential suppression 
of primordial power is a better fit  
to WMAP data than standard (unsuppressed) $\lcdm$.\footnote{
Of course this should not be surprising: 
the model for primordial power spectrum defined by Eq.~\ref{eq:pksupp} has been specifically constructed to improve 
the likelihood at large angles.}
In addition to this phenomenological motivation, Eq.~\ref{eq:pksupp} has been explored previously in the forecasting literature (in particular, Refs.~\cite{contaldi_etal03, mortonson_hu09,gibelyou10}) and so adopting this model facilitates a comparison of our calculations with existing results. 

The convention in the literature that has arisen to 
describe the large-angle anomaly is the so-called $\sonehalf$ statistic, defined as 
\beq
\label{eq:sonehalfdef}
\sonehalf\equiv\int_{-1}^{1/2}\dd(\cos \theta)\mathcal{C(\theta)}.
\eeq
As shown in Ref.~\cite{gibelyou10}, the value of the log of the cutoff parameter favored by the $\sonehalf$ statistic alone is 
$\logtenkc=-2.7,$ whereas a joint fit of both the $C_{\ell}$'s and $\sonehalf$ favors models with $\logtenkc=-3.3.$ Of course the likelihood of a given suppression model depends on the statistic used in the quantification, but regardless of this choice the likelihood is most sensitive to the cutoff scale, $k_c,$ so unless otherwise stated we choose parameter values $\alpha=3$ and $\beta=1$
 as our fiducial model of suppression and treat $k_c$ as a free parameter.

\subsection{Power Spectra}
\label{sub:powerspectra}

To assess the detectability of primordial power suppression in an imaging survey, our basic observables will be two-dimensional projected power spectra. 
The power spectrum $\mathcal{P}^{s_{i}s_{j}}(k,z)$ associated with the correlation function of a pair of three-dimensional scalar fields, $\ssubi$ and $\ssubj,$ can be related to 
its two-dimensional projected power spectrum, ${P}^{\xsubi\xsubj}(\ell),$ via the Limber approximation: 

\beq
\label{eq:limber}
P^{\xsubi\xsubj}(\ell)=\int dz \frac{\wiz\wjz}{\dasqz H(z)}\mathcal{P}^{\ssubi\ssubj}(k=\ell/\da(z),z).
\eeq
In Eq.~\ref{eq:limber}, valid for $\ell\gtrsim10,$ $\xsubi$ and $\xsubj$ are the 2-D fields that we observe as projections of
the 3-D fields, $\ssubi$ and $\ssubj,$ respectively. 
The angular diameter distance function is denoted by $\da,$ and $H(z)$ is the Hubble expansion parameter. The 2-D projected fields are related to the 3-D source fields through an integral over an appropriate weight function, $\wiz,$ associated with the observable of interest:

\beq
\label{eq:kernel}
\xsubi(\mathbf{\hat{\mathbf{n}}})=\int dz\wiz \ssubi(\mathbf{\hat{\mathbf{n}}},z).
\eeq
For galaxy fluctuations, the weight function is simply the redshift distribution\footnote{
The redshift distribution of galaxies, 
denoted as $n(\theta,\phi,z),$
is not to be confused with the three-dimensional
number density of galaxies, $n_{\mathrm{V}}(\vec{\mathbf{q}},\chi(z)),$
with $\chi$ the comoving radial coordinate and $\vec{\mathbf{q}}$ the transverse (2-D) coordinate.
The two distributions are simply related to each other as 
$n(\theta,\phi,z)\dd\Omega\dd z = n_{\mathrm{V}}\dd q^{2}\dd\chi,$
or $n(\theta,\phi,z)=\frac{\dasqz}{H(z)}n_{\mathrm{V}}(\vec{\mathbf{q}},\chi).$
}
of galaxies in the $i^{\mathrm{th}}$ tomographic bin, $\niz,$ times the Hubble expansion parameter:
\beq
\label{eq:galweight}
\mathrm{W}^{g}_{\mathrm{i}}(z)=H(z)\niz.
\eeq
For fluctuations in cosmic shear (convergence), the weight function is given by
\beq
\label{eq:shearweight}
W^{\kappa}_{\mathrm{i}}(z)=\frac{3}{2}H_{0}^{2}(1+z)\Omega_{\mathrm{m}}\daz\int_{z}^{\infty} dz'\frac{\da(z,z')}{\da(z')}n_{\mathrm{i}}(z'),
\eeq
where $\da(z,z')$ is the angular diameter distance between $z$ and $z'.$

In Eq.~\ref{eq:limber}, the three-dimensional power spectrum of the scalar field sourcing cosmic shear
is the matter power spectrum $\mathcal{P}^{\ssubi\ssubj}(k,z)=\mathcal{P}^{\deltasubi\deltasubj}(k,z),$ 
whereas for galaxy-galaxy correlations the source power is the 3-D galaxy power spectrum
$\mathcal{P}^{\ssubi\ssubj}(k,z)=\mathcal{P}^{\gi\gj}(k,z).$
Since we will be interested in the galaxy distribution on very large scales, it will suffice
for our purposes to relate the galaxy overdensity to the matter overdensity through a simple linear bias,
$\gi=b_{\mathrm{i}}\deltasubi,$ so that the galaxy power spectrum is related to the matter spectrum as 
$\mathcal{P}^{\gi\gj}(k,z)= b_{\mathrm{i}} b_{\mathrm{j}} \mathcal{P}^{\deltasubi\deltasubj}(k,z).$  
In all our calculations we use an independent galaxy bias function in each tomographic bin and allow each bias function to vary freely about a fiducial value of $b_{\mathrm{i}}(z)=1.$  
As the bias of most galaxy samples typically increases with redshift this simple prescription is nominally conservative, although in practice we find that our results concerning power suppression are insensitive to the fiducial bias function as well as the number of galaxy bias functions: the influence of scale-independent galaxy bias on our observables does not resemble the effect of a cutoff in the primordial power spectrum.

Above and throughout, lower-case Latin indices label the tomographic redshift bin of the sources.
In principle, the redshift distribution of the galaxies used for the galaxy power spectra need not be the same as that used for cosmic shear, but for simplicity we use the same underlying distribution
 for both so that the chief difference between the galaxy-galaxy power spectrum $\pgigj,$
the shear-shear power spectrum $\pkikj,$ and the cross-spectrum $\pkigj,$ is the form of the weight functions.  

As the models of primordial power suppression we study primarily affect very large scales, 
we will need to relax the Limber 
approximation in order to accurately predict the power at low $\ell.$  In this case, 
\beq
\label{eq:nolimber}
P^{\xsubi\xsubj}(\ell) = \frac{2}{\pi}\int\dd(\ln k)k^{3}P^{\deltasubi\deltasubj}(k)
\mathcal{I}_{\xsubi}(k)\mathcal{I}_{\xsubj}(k),
\eeq
where
$$\mathcal{I}_{\gi}(k)=\int\dd z \frac{\mathrm{W}^{g}_{\mathrm{i}}(z)}{H(z)}b_{\mathrm{i}}(z)\mathrm{D_{m}}(z)j_{\ell}(k\chi(z)),$$ 
$$\mathcal{I}_{\kappai}(k) = \int\dd z\frac{\mathrm{W}^{\kappa}_{\mathrm{i}}(z)}{H(z)}\mathrm{D_{m}}(z)j_{\ell}(k\chi(z)),$$
with $\mathrm{D_{m}}(z)\equiv\delta_{\mathrm{m}}(z)/\delta_{\mathrm{m}}(z=0)$ denoting the growth function, 
$\chi(z)$ the comoving distance, $b_{\mathrm{i}}$ the linear galaxy bias parameter, 
and $j_{\ell}(x)$ the spherical Bessel functions.  
See Ref.~\cite{loverde_afshordi_2008} for a recent, rigorous derivation of Eq.~\ref{eq:nolimber}.    

We model the underlying redshift distribution as 
$n(z)\propto z^{2}\exp{-(z/z_{0})}$ where the normalization is fixed so that $\int_{0}^{\infty}\dd z n(z)=\na,$ 
the surface density of sources in the survey.  As we will chiefly be interested in predictions for a very-wide-area photometric survey such as LSST or Euclid, 
unless explicitly stated otherwise we set $\na=30$  $\mathrm{gal/arcmin^2}$ and $z_0=0.34$ so that the median redshift is unity.  
We follow the treatment in Ref.~\cite{ma_etal06} and relate the tomographically binned 
galaxy distributions $\niz$ to the underlying redshift distribution 
according to $$\niz=n(z)\int_{z^{i}_{\mathrm{low}}}^{z^{i}_{\mathrm{high}}}\dd\zphot\phzpdf,$$ where 
${z^{i}_{\mathrm{low}}}$ and ${z^{i}_{\mathrm{high}}}$ are the boundaries of the $i^{\mathrm{th}}$ tomographic redshift bin.  
Photo-z uncertainty is controlled by the function $\phzpdf,$ a Gaussian at each redshift:
\beq
\label{eq:photozpdf}
\phzpdf = \frac{1}{\sqrt{2\pi}\sigma_{z}}\exp{\left[- \frac{(z - \zphot - \zbias)^{2}}{2\sigma_{z}^{2}}\right]}.
\eeq
The quantities $\sigma_z$ and $\zbias$ are themselves functions of redshift; we model the evolution of the spread 
as $\sigma_z = 0.05(1+z)$ and set $\zbias=0$ at all redshifts.

For a survey with its galaxies divided into $N_g$ redshift bins used to measure the galaxy correlation function, 
and $N_s$ bins for the galaxies used to measure cosmic shear,
there will be $N_g(N_g+1)/2$ distinct 2-D galaxy-galaxy power spectra $\pgigj,$ 
$N_s(N_s+1)/2$ distinct shear-shear power spectra $\pkikj,$ 
and $N_{s}N_{g}$ distinct cross-spectra $\pkigj.$

\subsection{Covariance}
\label{sub:covariance}

The covariance between a pair of observables, $P^{\xsubi\xsubj}$ and $P^{\xsubm\xsubn},$
is quantified by the covariance matrix
\beq
\label{eq:covone}
\mathrm{\mathbf{C}}[P^{\xsubi\xsubj},P^{\xsubm\xsubn}](\ell)
=\tilde{P}^{\xsubi\xsubm}(\ell)\tilde{P}^{\xsubj\xsubn}(\ell) + \tilde{P}^{\xsubi\xsubn}(\ell)\tilde{P}^{\xsubj\xsubm}(\ell).
\eeq
For the case of either galaxy-galaxy or shear-shear,
 the observed power spectra $\tilde{P}^{\xsubi\xsubj}$ have contributions from signal and shot noise,
 $$\tilde{P}^{\xsubi\xsubj}(\ell)=P^{\xsubi\xsubj}(\ell) + \mathrm{N}^{\xsubi\xsubj},$$
where $\mathrm{N}^{\gi\gj}=\delta_{\mathrm{ij}}N_{\mathrm{i}}^{\mathrm{A}}$ 
is the shot noise term for galaxy-galaxy spectra,
with $N_{\mathrm{i}}^{\mathrm{A}}$ denoting the surface density of sources, 
and $\mathrm{N}^{\kappai\kappaj}=\delta_{\mathrm{ij}}\gamma_{\mathrm{int}}^{2} N_{\mathrm{i}}^{\mathrm{A}}$ is the shear-shear shot noise term. We calculate the observed cross-spectra $\tilde{P}^{\kappai\gj}$ without a contribution from shot noise,
so that $\tilde{P}^{\kappai\gj}=P^{\kappai\gj}.$ 
We follow convention and set the intrinsic galaxy shape noise $\gamma_{\mathrm{int}}^{2}=0.2$
and absorb differences in shape noise between different observations into the surface density of sources.

\subsection{Forecasting}
\label{sub:forecasting}

We quantify the detectability of a primordial power suppression model
with a set of two-dimensional power spectra $P^{\xsubi\xsubj}$ in two distinct ways.  
First, we compute the $\chi^2$ difference between 
suppressed and unsuppressed power spectra:

\beq
\label{eq:chisqdef}
\Delta\chi^2=
\sum_{\mathrm{i},\mathrm{j},\mathrm{m},\mathrm{n},\ell}
\fsky(2\ell+1)\Delta\pxixj(\ell)\mathbf{C}^{-1}[\pxixj,\pxmxn](\ell)\Delta\pxmxn(\ell),
\eeq
where $\mathbf{C}^{-1}[\pxixj,\pxmxn](\ell)$ is the inverse of the covariance matrix   
associated with the {\em suppressed} power spectra at multipole $\ell,$ 
and the difference between suppressed and unsuppressed power spectra is denoted by 
$\Delta P=P_{\mathrm{sup}}-P_{\mathrm{unsup}}.$ 
In Eq.~\ref{eq:chisqdef}, $\fsky$ denotes the fraction of the sky covered by the survey.
The quantity $\sqrt{\Delta\chi^2}$ then represents the observable difference
 between the suppressed and unsuppressed models in units of the statistical uncertainty of the survey. 
 That is, if the observed power spectrum matches the unsuppressed ($\lcdm$) power spectrum 
 then the suppressed model could be ruled out at a significance of $\sqrt{\Delta\chi^2}$ sigmas. 
 
 Second, we employ the Fisher matrix to estimate the statistical constraints that a future survey
 will be able to place on the cutoff parameter $k_c$.  The Fisher matrix is defined as 
\beq
\label{eq:fisher}
F_{\alpha \beta}=\sum_{\ell=2}^{\ell_{\mathrm{max}}} (2\ell+1)\fsky 
\sum_{\mathrm{i,j,m,n}} \frac{ \partial \pxixj }{\partial p_{\alpha}} 
\mathbf{C}^{-1}[\pxixj,\pxmxn]
\frac{ \partial \pxmxn }
{\partial p_{\beta}} + F_{\alpha \beta}^{\mathrm{P}}. 
\eeq
The parameters of the model are $p_{\alpha}$ and $p_{\beta},$ with Greek indices labeling the model parameters.
For all our observables we set $\ell_{\mathrm{max}}=300;$ this ensures that the assumptions of weak lensing and 
Gaussian statistics are valid \cite{white_hu00,cooray_hu01,vale_white03,dodelson_etal06,semboloni_etal06} 
and that modeling the galaxy distribution with a simple linear bias is appropriate; 
most interesting models of power suppression are constrained by multipoles 
$\ell<30,$ firmly in the linear galaxy bias regime, so we expect that our scale-independent bias assumption is well-founded.

The inverse of the Fisher matrix is an estimate of the parameter covariance near 
the maximum of the likelihood, i.e.~at the fiducial values of the parameters.  
The measurement error on parameter $\alpha$ marginalized 
over all other parameters is
%
\beq
\label{eq:ferror}
\sigma(p_{\alpha})=[F^{-1}]_{\mathrm{\alpha \alpha}}.
\eeq
Throughout this work we assume flat spatial geometry and allow seven cosmological parameters to vary about the following fiducial values:
$\Omegam h^2= 0.13,$  $\wzero=-1,$ $\wa=0.0,$ $\Omegab h^2 = 0.0223,$ $\ns=0.96,$ 
$\mathrm{ln}(\dr)=-19.953,$ and $\Omega_{\Lambda}=0.73.$

Gaussian priors on the parameters are incorporated into the Fisher analysis via the second term
in Eq.~\ref{eq:fisher}. In all our forecasts we use the following priors: 
$\Delta\Omegam h^2 = 0.007,$ $\Delta\Omegab h^2 = 0.001,$ $\Delta\ns = 0.04,$ $\Delta\mathrm{ln}(\dr)=0.1;$
these are comparable to current marginalized constraints on these parameters \cite{komatsu_etal11} and so represent a 
conservative estimate for the statistical uncertainty on cosmology that will be achieved 
in advance of LSST or Euclid.

\section{Results}
\label{section:results}

\subsection{Effect of Suppression on Power Spectra}
\label{sub:pertspectra}

Figure \ref{fig:fracdiff} illustrates the effect of a particularly aggressive model of power suppression on the observables.  
The fractional change to $P^{\kappa\kappa}$ (dashed curves), 
$P^{gg}$ (solid curves), and $P^{\kappa g}$ (dot-dashed curves) are each plotted as a function of multipole.
In calculating the power spectra plotted in Fig.~\ref{fig:fracdiff}, we have
used four tomographic bins, evenly spaced in the range $0<z<3,$
 for the distribution of both galaxy correlation sources and shear sources, so that the
thin blue curves trace the change to each signal in the tomographic bin with redshift boundaries $0.75<z<1.5,$ 
and the thick red curves trace changes in the bin with redshift boundaries $1.5<z<2.25.$ 

\FIGURE[t!]{\epsfig{file=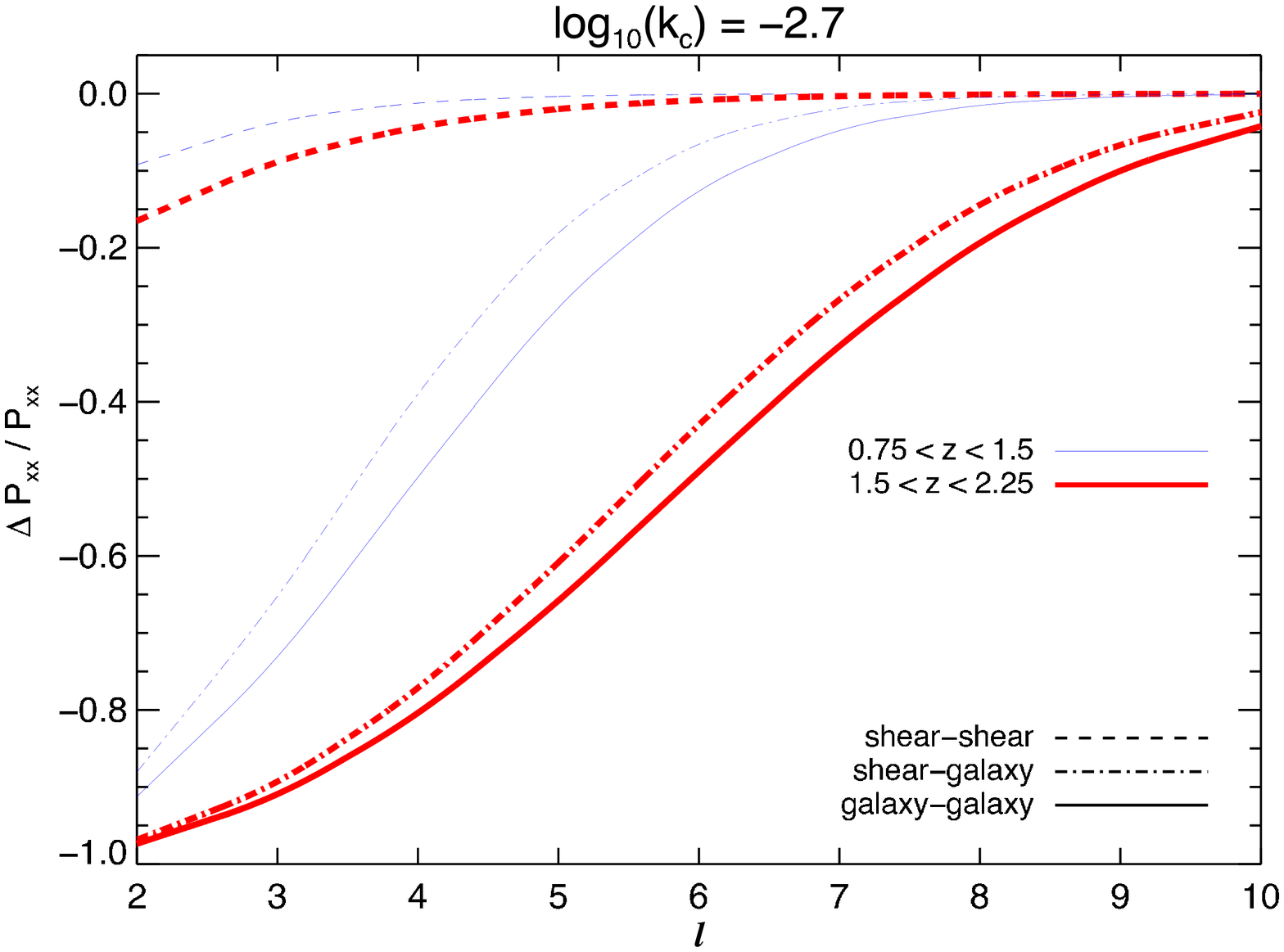,width=12cm}
\caption{
Toy demonstration of the effect of power suppression on the three sets of power spectra we study:
galaxy-galaxy are plotted with solid curves, shear-shear with dashed, and shear-galaxy with dot-dashed.  
The fractional change to each observable,
 $(P^{\xsubi\xsubj}_{\rm sup}-P^{\xsubi\xsubj}_{\rm unsup}) / P^{\xsubi\xsubj}_{\rm unsup},$  
 is plotted as a function of multipole.
 The fractional change to the signal at high redshift is plotted with thick, red lines for each observable;
 the signal at low redshift with thin, blue lines.  
High-redshift bins are more fractionally perturbed than low because for a fixed angular scale,
larger redshift corresponds to larger physical scale.  
Galaxy-galaxy power correlations are more affected than shear-shear since the redshift
kernel peaks at higher redshift for galaxy-galaxy (the lenses are in front of the galaxies).
}
\label{fig:fracdiff}
}

Two features of this figure are particularly worthy of note:
\ben
\item The signal at high redshift is more sensitive to large-scale primordial power suppression
than the signal at low redshift, irrespective of the observable.
\item Galaxy-galaxy power spectra are more sensitive than shear-shear.
\een

Both of these features are simple to understand.  For a fixed angular scale,
larger redshift corresponds to larger physical scale, so that for $i>j,$ $P^{\xsubi\xsubi}$
probes the matter distribution on larger scales than $P^{\xsubj\xsubj}$ and thus $P^{\xsubi\xsubi}$ will be more 
dramatically affected by large-scale power suppression.

\FIGURE[t!]{\epsfig{file=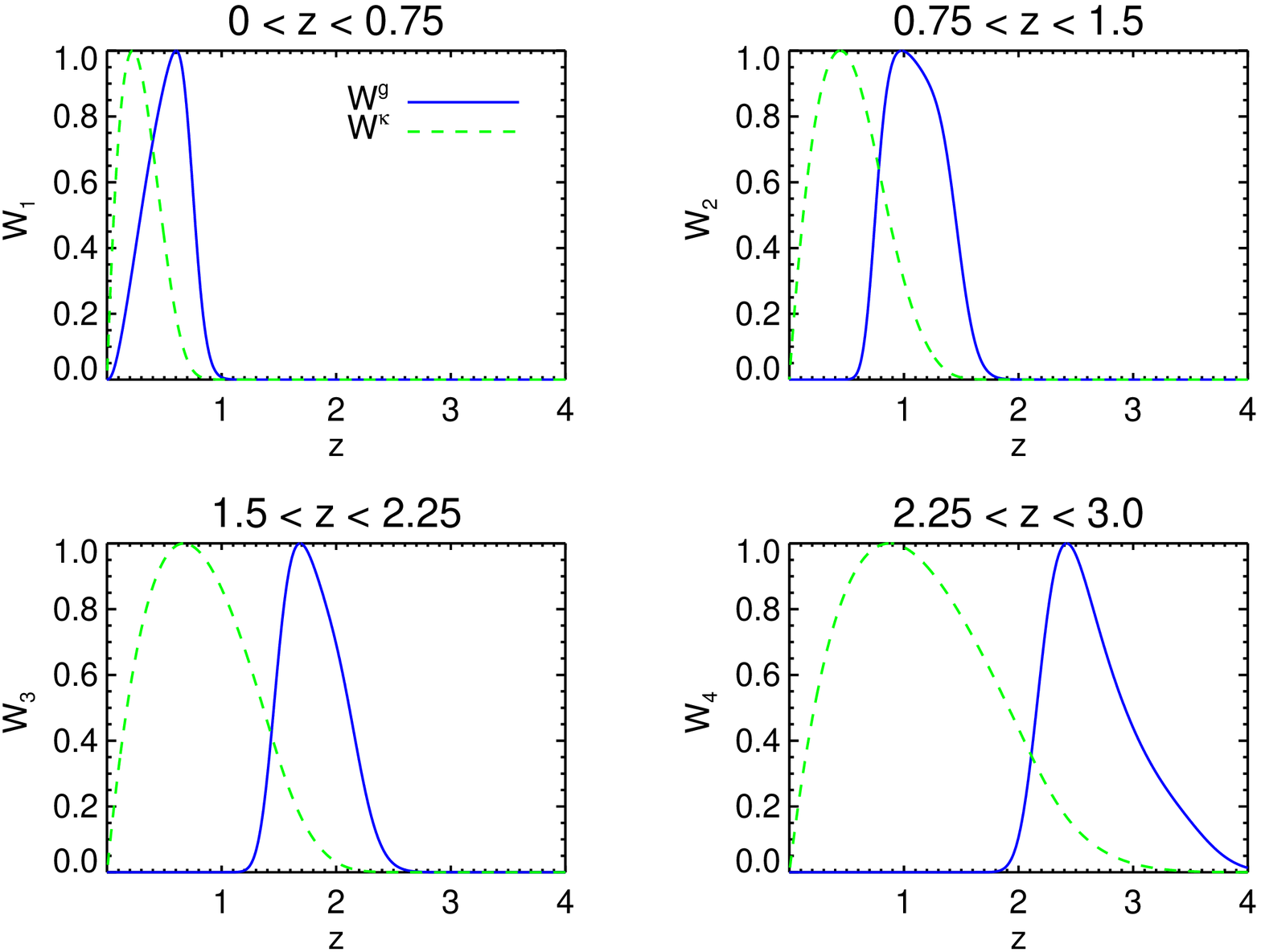,width=12cm}\caption{
Comparison of the weight functions (Eqns.~\ref{eq:galweight} \& \ref{eq:shearweight}) for a toy set of 
galaxy-galaxy auto-power spectra (solid blue curves) and 
shear-shear auto-power spectra (dashed green curves), each divided into four tomographic bins. 
 Each kernel has been normalized by its own
maximum value to facilitate a direct comparison of the redshift evolution of the weight functions.
}
\label{fig:weights}
}

Galaxy-galaxy power spectra are more sensitive to large-scale features in $\pmat(k)$ than shear-shear 
for a related reason.
  In Fig.~\ref{fig:weights} we compare the weight functions (defined by Eqns.~\ref{eq:galweight} and \ref{eq:shearweight}) in each of four tomographic bins  
for $P^{\kappa\kappa}$ (dashed green) and $P^{gg}$ (solid blue). 
Since $\mathrm{W}_{\mathrm{i}}^{g}$ peaks at larger redshift than 
 $\mathrm{W}_{\mathrm{i}}^{\kappa},$  $P^{gg}$ is comparatively more sensitive to physics at high redshift
 than $P^{\kappa\kappa}.$  This is simply because the galaxy-galaxy auto-power spectrum
  in a particular redshift bin probes the clustering properties of matter {\em within} that bin, whereas 
  the shear-shear auto-power spectrum probes the distribution of matter that lies {\em in between} the source galaxies
  and the telescope.  Thus, as discussed above, for a fixed angular scale 
  $P^{gg}$ is comparatively more sensitive to gravitational clustering on large scales than $P^{\kappa\kappa},$
  and therefore large-scale power suppression will induce a greater fractional change in $P^{gg}$
  than $P^{\kappa\kappa}.$ 

\subsection{Detectability}
\label{sub:chiresult}

In this section we employ the 
$\Delta\chi^2$ technique described in \S\ref{sub:forecasting} to study the ability of a very-wide-area photometric survey to distinguish between 
power spectra in standard $\lcdm$ and a model in which primordial power is suppressed on scales 
$k\lesssim k_{c}$ according to Eq.~\ref{eq:pksupp}.  
Our chief result for this technique is plotted in Figure \ref{fig:chivskc}. 
The square root of $\Delta\chi^2,$ computed via Eq.~\ref{eq:chisqdef}, appears on the vertical axis;
the log of the cutoff scale appears on the horizontal axis.  
We have used the suppressed-model power spectra to calculate the covariance matrix in Eq.~\ref{eq:chisqdef}, 
so that Fig.~\ref{fig:chivskc} is suited to answer the following question:
if the true model of the primordial power spectrum is unsuppressed, so that a LSST- or Euclid-like survey observes the large scale galaxy clustering statistics and cosmic shear signal predicted by standard $\lcdm,$ then as a result of such observations to what confidence could we rule out a given model of power suppression (i.e.~a given value of $k_c$)? 
Thus the vertical axis values in Fig.~\ref{fig:chivskc} represent the confidence (the ``number of sigmas") 
with which the suppressed model can be ruled out by planned imaging surveys.  
With the dotted magenta line we have plotted the 
detectability of power suppression with a spectroscopic survey similar to BigBOSS, 
corresponding to $\fsky=0.5$ and $\na=0.5$ $\galarcmintwo;$  
we have used the same galaxy distribution for our BigBOSS calculations as the $n(z)$ we used for our imaging survey  
to facilitate a direct comparison between these two surveys.  
The technique used for the calculation of $\Delta\chi^2$ for the case of a 
spectroscopic survey is straightforward; 
 we refer the reader to Ref.~\cite{gibelyou10} for details. The detectability of power suppression with an LSST- or Euclid-like survey using $P^{\kappa\kappa}$ is shown by the dashed green curve, 
using $P^{gg}$ the dot-dashed blue curve, and a joint analysis the solid red curve; detectability levels of $1\sigma$ and $3\sigma$ are delineated by solid, gray horizontal lines.  There is little added advantage that a joint analysis has over 
using the galaxy-galaxy power spectra alone because $P^{gg}$ is much more sensitive to horizon-size scales 
than $P^{\kappa\kappa}$ (see the discussion in \S\ref{sub:forecasting}).

Of particular interest here is the comparison between the dotted and solid curves, 
which can be thought of as contrasting the constraining power of the most ambitious large-scale 
spectroscopic and photometric surveys that will be undertaken over the next ten years.   
In this context, the advantage a spectroscopic redshift survey has over a photometric imaging survey 
is that the redshift survey has many more radial modes available to probe very large scales: 
photometric redshift uncertainty restricts the sampling of the radial signal to a small handful of 
tomographic redshift bins.  
Thus in general, the constraining power of an imaging survey increases with the number of tomographic bins as finer binning allows 
for more detailed study of the redshift evolution of the signal.  This information eventually saturates and further refinement of the binning ceases to significantly  improve the constraints.  We find very little improvement in the $\Delta\chi^2$ results beyond $N_s=4$ tomographic bins used for cosmic shear and $N_g=8$ redshift bins for galaxy correlations.  
The limitation to the radial information that necessarily comes from using photometric data is more than compensated for by the greater surface density of sources that will be observed by LSST or Euclid.  This may seem surprising since power suppression primarily affects multipoles $\ell\lesssim30$ where cosmic variance is typically thought to dominate the errors.  However, because most of the constraining power on primordial power suppression comes from sources at high redshift where the surface density of galaxies is quite sparse, 
shot noise is significant and imaging surveys will be able to exploit this relative advantage to distinguish at the $3\sigma$ level between $\lcdm$ and models of power suppression 
that are favored by the $\sonehalf$ statistic, $\logtenkc\gtrsim-2.7.$
Models that are mutually favored by both the CMB $C_{\ell}$'s and $\sonehalf,$ 
$\logtenkc\lesssim-3.3,$ will not produce an effect that will be statistically significant in data sets that will be obtained by LSST, Euclid, or BigBOSS, 
  and hence these models will remain inaccessible to the galaxy surveys currently planned to take place within the next decade.

\FIGURE[t!]{\epsfig{file=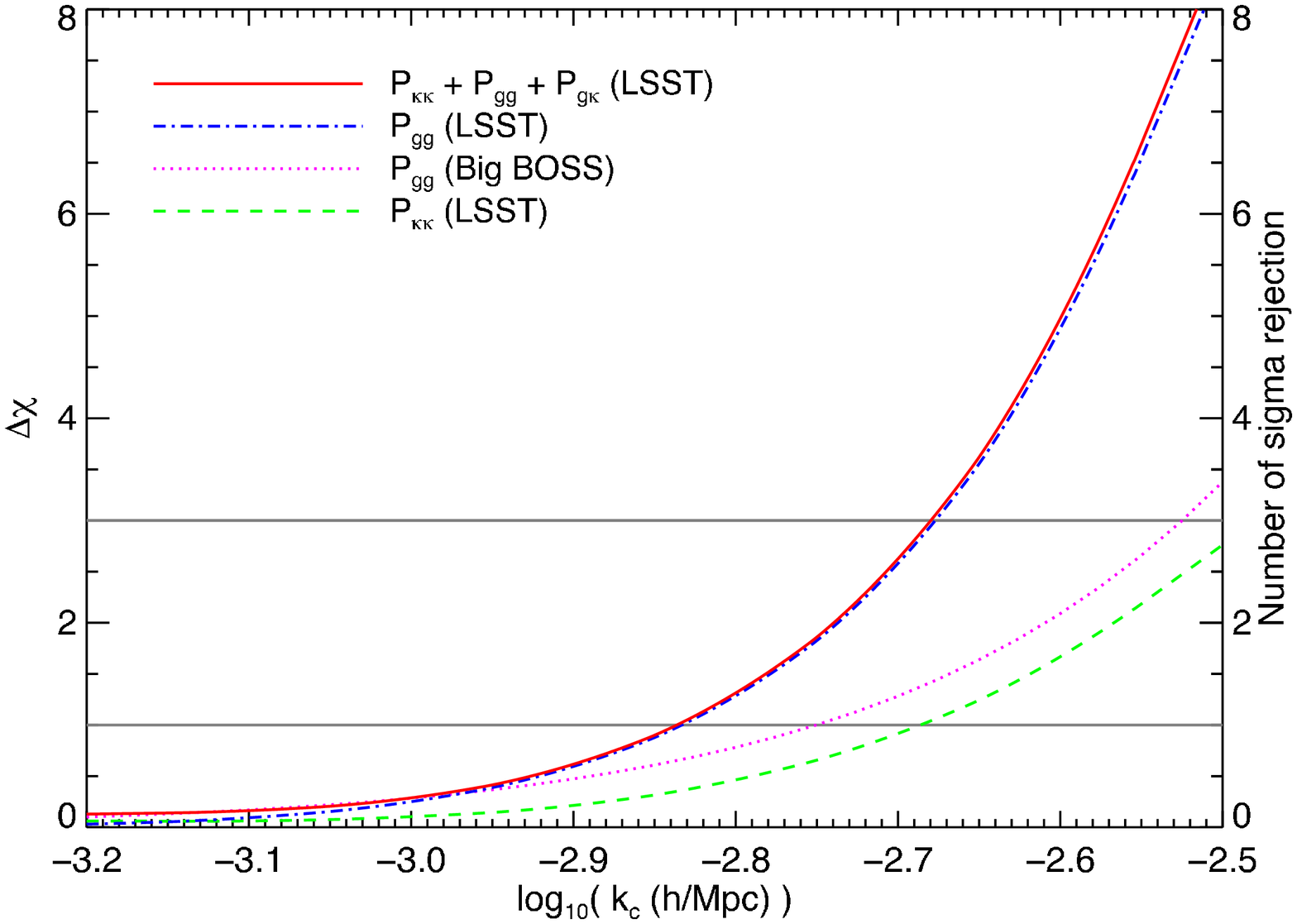,width=12cm}\caption{
The cutoff scale at which primordial power suppression becomes significant appears on the horizontal axis. Along the vertical axis is $\Delta\chi=\sqrt{\Delta\chi^2}$, defined by Eq.~\ref{eq:chisqdef}, our first statistic quantifying the detectability of power suppression with future wide-area surveys. 
As discussed in \S~\ref{sub:pertspectra}, galaxy clustering has significantly greater constraining power on primordial power suppression models than cosmic shear. Models of power suppression favored by the $\sonehalf$ statistic alone, $\logtenkc=-2.7,$ could be ruled out at the $3\sigma$ level by a joint analysis of an LSST- or Euclid-like survey; models with $\logtenkc=-3.3$ that are mutually favored by the $C_{\ell}$'s and $\sonehalf$ will be inaccessible to the surveys we consider.
}
\label{fig:chivskc}
}

\subsection{Statistical Constraints on $k_c$}
\label{sub:sigmaresult}

In this section we present our results for the second method we used to assess the detectability of primordial power 
suppression, in which we employ the Fisher matrix formalism to forecast statistical constraints on the cutoff parameter $k_c.$ The advantage this approach has over the method described in \S\ref{sub:chiresult} is that the Fisher 
formalism provides a natural way to account for degeneracies between $k_c,$ cosmological parameters, 
and galaxy bias, as well as a way to estimate the significance of photo-z uncertainty.  
We use seven cosmological parameters in our Fisher analysis with the same fiducial values and priors specified in 
\S\ref{sub:forecasting}. 

The results from this calculation appear in Figure \ref{fig:sigmavskc}.  On the horizontal axis is the log of the 
cutoff scale, on the vertical axis the statistical constraint on $\logten(k_c).$ 
Constraints on $\logten(k_c)$ when only convergence power spectra are used appear as the dashed green curve, 
using only $P^{gg}$ as the dot-dashed blue curve, and a joint analysis as the solid red curve.  
Just as we found for our $\Delta\chi^2$ results in \S\ref{sub:chiresult}, 
the constraining power on the cutoff saturates at $N_g=8$ tomographic redshift bins 
for galaxy power spectra and $N_s=4$ redshift bins for cosmic shear.

The results presented in Fig.~\ref{fig:sigmavskc} are suited to answer the following question: 
if the true primordial power spectrum is, in fact, exponentially suppressed beyond some scale $k_c$, 
so that a LSST- or Euclid-like survey observes the large scale galaxy clustering statistics and cosmic shear signal predicted by the suppressed model, then to what statistical precision could the parameter $k_c$ be constrained by such an observation?
Thus this calculation is complementary to the one presented in the previous section in the following sense: 
results in \S\ref{sub:chiresult} pertain to the difference between a given power suppression model and $\lcdm,$  
whereas results in this section pertain to the difference between one power suppression model and another (nearby in $k_c$ parameter space) power suppression model.
The salient conclusions that can be drawn from Fig.~\ref{fig:sigmavskc} are similar to those in Fig.~\ref{fig:chivskc}; 
when $\logtenkc=-2.7$ a joint analysis provides a relatively tight $7\%$ constraint on the cutoff parameter; 
 for power suppression models that are mutually favored by both the $\sonehalf$ statistic and the CMB multipoles
 the cutoff is on larger scales, $\logtenkc=-3.3$, where the constraining power of a survey such as LSST or Euclid is comparably weak.

In order to perform these calculations, the derivatives appearing in Eq.~\ref{eq:fisher} must be evaluated numerically; we found that our results are robust to changes in both step-size as well as the choice to compute one- or two-sided derivatives for all parameters in the analysis. In particular, the statistical constraints on $k_c$ are in principle asymmetric because constraining power varies monotonically with the scale of the cutoff. However, we find that the constraints vary sufficiently slowly with $k_c$ to limit the effect of the asymmetry to a correction of only a few percent over the relevant range of parameter space. 
 
Within reasonable levels, photometric redshift uncertainty turns out to have very little effect 
on the results in Fig.~\ref{fig:sigmavskc}.  We have checked this conclusion in two distinct ways.  
First, we parametrize photo-z uncertainty as described in Ref.~\cite{ma_etal06}.  
Briefly, the redshift evolution of each of the functions $\sigma_z$ and $\zbias$ is modeled by linearly interpolating 
among a set of 31 control points, one at each interval of $\delta z=0.1$ between $z=0$ and $z=3$, where the values 
at the control points are given by $\sigma_z=A(1+z)$ and $\zbias=0,$ with $A=0.05$ as our standard value 
of the photo-z spread at redshift zero. 
These $31\times2=62$ control points then serve as photo-z uncertainty parameters in the Fisher analysis.
We find that the power spectrum cutoff parameters exhibit very little degeneracy with photometric redshift parameters, so that the constraints are not degraded significantly by marginalizing over these or more complex photometric redshift parameterizations.
Second, we studied the sensitivity of the constraints to the value of $A=\sigma_z/(1+z).$  
The chief effect that varying $A$ has on our results comes from restricting the statistically independent information 
in the tomography, i.e.~larger values of $A$ lead to constraining power that saturates at smaller numbers of 
tomographic redshift bins because photo-z uncertainty smears out correlations along the line of sight and thus restricts 
the number of radial modes available to probe cosmology.  However, for reasonable values of $A$ this 
effect is quite small: $\sigma(\logten(k_c))$ only changes by roughly ten percent when $A$ varies between the optimistic value of $A=0.03$ and the quite pessimistic value of $A=0.15.$  As the effect of photo-z uncertainty 
on the dark energy constraints is much more profound \cite{ma_etal06,detf,huterer_etal06,lima_hu07,zentner_bhattacharya09,bernstein_huterer09,zhang_etal09,hearin_etal10}, the photo-z calibration effort leading up to future wide-area imaging surveys 
will very likely achieve the precision required to reach the constraining power illustrated in 
Figs.~\ref{fig:chivskc}-\ref{fig:sigmavskc}.

\FIGURE[t!]{\epsfig{file=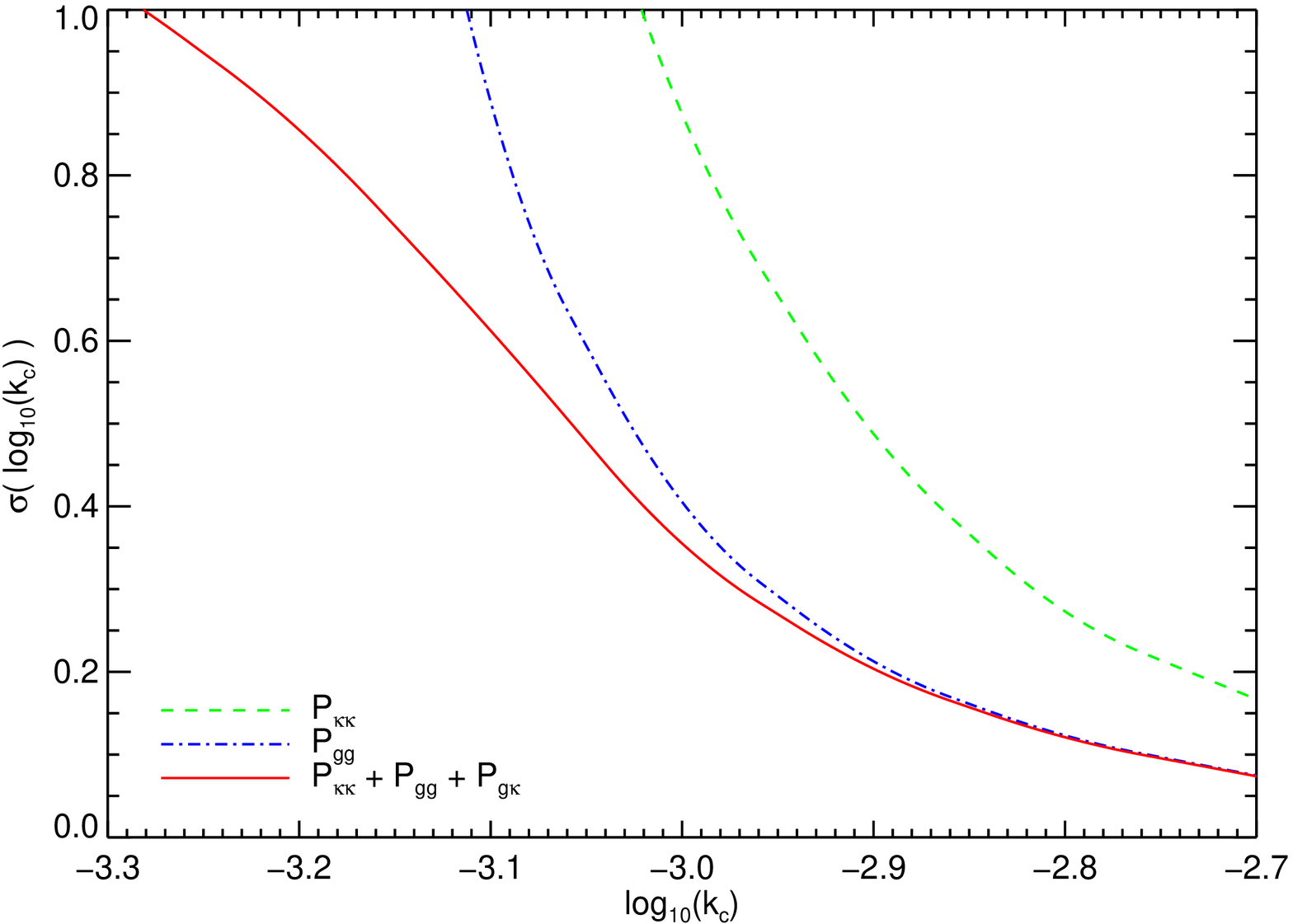,width=12cm}\caption{
Plot of the statistical constraining power that future imaging surveys will have on the cutoff scale.  We compute 
$\sigma(\mathrm{log_{10}}(k_c)),$ the precision to which the scale of the cutoff can be measured, using the Fisher 
matrix formalism; forecasts for the constraints from an LSST- or Euclid-like imaging survey using 
galaxy clustering statistics alone appear as the dot-dashed blue curve, using cosmic shear alone as the dashed green curve, 
and with a joint analysis the solid red curve.
}
\label{fig:sigmavskc}
}

\section{Conclusions and Discussion}
\label{section:discussion}

We have studied the sensitivity of an all-sky galaxy imaging survey such as LSST or Euclid to the suppression of primordial 
power on scales comparable to the horizon size at the time of recombination.  The models of suppression we 
investigated are motivated by the observed deficit in the two-point correlation function of the CMB temperature at 
large angles.  In particular, we focused on constraining the comoving cutoff scale $k_c$ at which exponential 
suppression of $\Delta^{2}_{\delta}(k)$ sets in.

The cutoff scale favored by the $\sonehalf$ statistic alone is $\sim700~$Mpc.  We find that a LSST- or Euclid-like survey will be able to 
distinguish this model from $\lcdm$ at the $3\sigma$ level, and that if we do in fact live in a universe in which 
the primordial power spectrum is exponentially suppressed on comoving scales larger than $\sim700~$Mpc, 
then a joint analysis of shear and galaxy correlations could provide $7\%$ constraints 
on the cutoff parameter.  However, planned galaxy surveys will not be able to discriminate between $\lcdm$ and models 
in which the suppression sets in at cutoff scales $\sim2.8~$Gpc, as favored jointly by the 
$C_{\ell}$'s and $\sonehalf$. 
The chief reason for the relatively weak constraining power on models with a cutoff on these larger scales is simply that the redshift range covered by the next generation of galaxy surveys is not deep enough to probe matter-power-spectrum modes larger than a few Gpc. Thus it may be necessary to rely on future observations of the CMB, particularly the polarization signal and its cross-correlation with the temperature (as studied in Ref.~\cite{mortonson_hu09}), to test power suppression models on these very large scales.

In our forecasts of the constraints on power suppression we have not taken into account possible systematic errors that complicate any observation of galaxy clustering or cosmic shear on large scales.  Any calibration error that varies across the sky over a difference of $\sim20^{\circ}$ can interfere with a measurement of the large-scale clustering signal and therefore contribute to the error budget of a galaxy survey. For example, an effective change to the magnitude limit of a survey can be induced by Galactic extinction, which may vary with the line of sight through the Milky Way. The relative frequency of star-galaxy misclassifications may effectively vary across the sky due to the spatial dependence of stellar demographics in the Galaxy.  We leave a detailed assessment of possible sources of systematic error as a task for future work.

 Finally, we have compared the sensitivity to large-scale features in 
$\Delta^{2}_{\delta}(k)$ of a spectroscopic redshift survey with characteristics similar to that of BigBOSS and future wide-area imaging surveys such as LSST or Euclid.  We find that while a BigBOSS-like survey has 
a greater number of statistically independent radial modes with which to probe large scales, planned imaging surveys will be more sensitive to horizon-scale physics because of the greater surface density of high-redshift sources they will observe.

\acknowledgments
We thank Jeff Newman, Rupert Croft, Michael Wood-Vasey, Niayesh Afshordi, and Aravind Natarajan for helpful discussions.
We especially thank Dragan Huterer for providing useful feedback throughout the completion of this work.
APH and ARZ are supported by the University of Pittsburgh and by the National Science Foundation 
through grant AST 0806367. CMG is supported by NASA under contract NNX09AC89G.

\bibliography{supppap}

\end{document}